\documentclass[a4paper,11pt]{article}
\usepackage{pos}
\usepackage{here}
\usepackage{url}
\usepackage{graphicx}% Include figure files
\usepackage{dcolumn}% Align table columns on decimal point
\usepackage{bm}
\usepackage{comment}

\newcommand{\TTRR}{\mathbb{T}^2\times\mathbb{R}^2}
\newcommand{\SRRR}{\mathbb{S}^1\times\mathbb{R}^3}

\title{Anisotropic pressure and novel first-order phase transition in SU(3)
 Yang-Mills theory on $\mathbb{T}^2\times\mathbb{R}^2$}
\author*[a,b]{Daisuke Fujii}
\author[b]{Akihiro Iwanaka}
\author[c,d]{Masakiyo Kitazawa}
\author[e]{Daiki Suenaga}

\affiliation[a]{Advanced Science Research Center, Japan Atomic Energy Agency (JAEA), \\
Tokai, 319-1195, Japan}

\affiliation[b]{Research Center for Nuclear Physics, Osaka University, \\
Ibaraki 567-0048, Japan}

\affiliation[c]{Yukawa Institute for Theoretical Physics,
Kyoto University, \\
Kyoto 606-8502, Japan}

\affiliation[d]{J-PARC Branch, KEK Theory Center, 
  Institute of Particle and Nuclear Studies, KEK, \\ Tokai, Ibaraki 319-1106, Japan}

\affiliation[e]{Kobayashi-Maskawa Institute for the Origin of Particles and the Universe,
Nagoya University, \\
Nagoya, 464-8602, Japan}

\emailAdd{daisuke@rcnp.osaka-u.ac.jp}

\abstract{
We investigate the thermodynamic behavior and phase diagram of $SU(3)$ Yang-Mills theory on $\mathbb{T}^2 \times \mathbb{R}^2$ in Euclidean spacetime using an effective model. In our approach, the Polyakov loops along the compactified directions are treated as dynamic variables, and the model is calibrated to match lattice simulation results for thermodynamic observables on $\mathbb{T}^2 \times \mathbb{R}^2$. Our analysis reveals a novel first-order phase transition in the deconfined phase that ends at critical points, which appear to belong to the two-dimensional $Z_2$ universality class. This transition is driven by the interplay between the two Polyakov loops, introduced via a cross-term in the Polyakov-loop potential.}

\FullConference{The XVIth Quark Confinement and the Hadron Spectrum Conference (QCHSC24)\\
 19-24 August, 2024\\
 Cairns Convention Centre, Cairns, Queensland, Australia\\}

\begin{document}
\maketitle

\section{\label{sec:level1}introduction}

Thermodynamic quantities, such as pressure and energy density, are essential observables that characterize the properties of a medium in equilibrium. These observables play a pivotal role in uncovering the phase structure in Quantum Chromodynamics (QCD) and pure Yang-Mills (YM) theories. Although spatial isotropy is often assumed, thermodynamics can also be applied to systems with spatial anisotropy, which can be realized by imposing specific boundary conditions.

In Ref.~\cite{Kitazawa:2019otp}, the anisotropic thermodynamics of $SU(3)$ YM theory on $\mathbb{T}^2\times\mathbb{R}^2$ was investigated through lattice simulations for temperatures above the deconfined transition temperature $T_{\rm d}$. It was observed that the pressure anisotropy in this system is significantly suppressed compared to that in a free boson theory. This suppression is expected to stem from the nonperturbative nature of gluodynamics, and elucidating its origin may provide deeper insights into the nonperturbative dynamics of the theory.

In the present work~\cite{Fujii:2024llh}, we extend previous effective model approaches~\cite{Suenaga:2022rjk} by formulating a model on $\mathbb{T}^2\times\mathbb{R}^2$ that incorporates the interplay between two Polyakov loops associated with the temporal and spatial directions. By introducing cross terms in the potential to capture this interplay, our model qualitatively reproduces the lattice data for $T/T_{\rm d}\gtrsim1.5$ and predicts a novel first-order phase transition in the deconfined phase, with critical endpoints that likely belong to the two-dimensional $Z_2$ universality class. These findings offer a promising avenue for further exploration of anisotropic thermodynamics and the underlying nonperturbative phenomena in QCD and YM theory.

\section{\label{sec:level2}Formulations}

In this section we build the effective model of the $SU(3)$ YM theory on $\mathbb{T}^2\times\mathbb{R}^2$ by using two Polyakov loops. In this proceedings, we assume that the $\tau$ and $x$ axes are compactified by length $L_\tau$ and $L_x$ with periodic boundary conditions (PBCs), respectively, and that the $y$ and $z$ directions are extended infinitely.

\subsection{Polyakov loops}

For the construction of our effective Polyakov loop model, we first review the definition of the Polyakov loop. In $SU(3)$ Yang-Mills theory on $\mathbb{T}^2\times\mathbb{R}^2$, we define two Polyakov loops along the compactified directions as 
\begin{align}
    \Omega_c(\bm{x}_c^\perp)=\frac{1}{N}{\rm Tr}\Big(\hat\Omega_c(\bm{x}_c^\perp)\Big),\qquad  \left(\hat\Omega_c(\bm{x}_c^\perp)=\mathcal{P}\exp\Big(i\int^{L_c}_0A_c(x_c,\bm{x}_c^\perp)dx_c\Big)\right), 
    \label{eq:PolyakovLoopDef}
\end{align}
where $A_\mu(x)$ is the $SU(N)$ gauge field, $\bm{x}_\tau^\perp=(x,y,z)$, $\bm{x}_x^\perp=(\tau,y,z)$, and $\mathcal{P}$ denotes path ordering. 
For $N=3$, the gauge field is diagonalized by a gauge transformation, where the Polyakov loops are written as $\hat\Omega_c(\bm{x}_c^\perp) = {\rm diag}[e^{(\theta_c)_1},e^{(\theta_c)_2},e^{(\theta_c)_3}]$ with $\sum^3_{j=1}(\theta_c)_j=0 \ ({\rm mod} ~ 2\pi)$. 
Following Refs.~\cite{Meisinger:2001cq,Dumitru:2012fw}, we employ the ansatz $(\theta_c)_j=(2-j)\phi_c \ (j=1,2,3)$,
so that the Polyakov loop becomes
\begin{align}
    \Omega_c=\frac{1}{3}\sum_{j=1}^{3} \exp\bigl(i(\theta_c)_j\bigr)=\frac{1}{3}\Bigl(1+2\cos\phi_c\Bigr),
    \label{Pctheta}
\end{align}
which depends on a single parameter $\phi_c$. In the following, we assume $0\leq\phi_c\leq2\pi/3$. Then, $Z_3^{(c)}$ symmetric phase, i.e., $\Omega_c=0$ is realized in $\phi_c=2\pi/3$, while $\Omega_c=1$ is realized in $\phi_c=0$.

\subsection{Free energy}

To investigate the thermodynamics on $\mathbb{T}^2\times\mathbb{R}^2$, we employ an effective model based on two Polyakov loops~\cite{Suenaga:2022rjk,Fujii:2024llh}. In the original work~\cite{Meisinger:2001cq}, the isotropic thermodynamics of YM theory on $\mathbb{S}^1\times\mathbb{R}^3$ is described using a single Polyakov loop, $\Omega_\tau(\bm{x}_\tau^\perp)$. In this study, we extend this approach to incorporate spatial anisotropy by using the free energy that depends on both $L_\tau$ and $L_x$ constructed by $\Omega_\tau$ and $\Omega_x$. Following the methods of Refs.~\cite{Meisinger:2001cq,Dumitru:2012fw}, we write 
\begin{align}
    f(\vec\theta_\tau,\vec\theta_x;L_\tau,L_x)= f_{\rm pert}(\vec\theta_\tau,\vec\theta_x;L_\tau,L_x)
    + f_{\rm pot}(\vec\theta_\tau,\vec\theta_x;L_\tau,L_x),
    \label{eq:f}
\end{align}
with $\vec{\theta}_\tau$ and $\vec{\theta}_x$ determined by minimizing this free energy.  

According to Ref.~\cite{Sasaki:2012bi}, the one-loop free energy per unit volume with background field $A_\mu$ corresponding to the Eq.~\eqref{Pctheta}, i.e., the free energy of perturbative gluons, is given as follows: 
\begin{align}
&f_{\rm pert}(\vec{\theta}_\tau,\vec{\theta}_x;L_\tau,L_x) \notag \\
&= -\frac{8\pi^2}{45L^4_\tau}+\frac{8\phi^2_\tau(\phi_\tau-\pi)^2+\phi^2_\tau(\phi_\tau-2\pi)^2}{6\pi^2L^4_\tau}-\frac{8\pi^2}{45L^4_x}+\frac{8\phi^2_x(\phi_x-\pi)^2+\phi^2_x(\phi_x-2\pi)^2}{6\pi^2L^4_x} \notag \\
&\hspace{5mm}-\frac{8}{\pi^2}\sum^\infty_{l_\tau,l_x=1}\frac{1}{{X^4_{l_\tau,l_x}}}\Big[1+2\cos(\phi_\tau l_\tau)\cos(\phi_x l_x)+\cos(2\phi_\tau l_\tau)\cos(2\phi_x l_x)\Big],\label{fpert2}
\end{align}
with $\vec\theta_c=(\phi_c,0,-\phi_c)$ and $X_{l_\tau,l_x}\equiv\sqrt{(l_\tau L_\tau)^2+(l_xL_x)^2}$ in Refs.~\cite{Suenaga:2022rjk,Fujii:2024llh}.
In the limit $L_x\to\infty$, only the first two terms survive and Eq.~\eqref{fpert2} reduces to the one-loop free energy $f_{\rm pert}^{\SRRR}$ in Refs.~\cite{Meisinger:2001cq,Dumitru:2012fw,Sasaki:2012bi}. 
Considering only this term, the free energy \eqref{fpert2} always has a minimum at $\phi_c=0$ and the confined phase is never realized. 

\subsection{Potential term}

To achieve the confined phase, it is essential to include a potential term in our effective model. We decompose this potential into two parts:
\begin{align}
    f_{\rm pot}(\vec{\theta}_\tau,\vec{\theta}_x;L_\tau,L_x) = f_{\rm sep}(\vec{\theta}_\tau,\vec{\theta}_x;L_\tau,L_x) + f_{\rm cross}(\vec{\theta}_\tau,\vec{\theta}_x;L_\tau,L_x).
\end{align}
The form of the potential is determined by imposing general constraints on the free energy (see Refs.~\cite{Suenaga:2022rjk,Fujii:2024llh} for details). For instance, the theory must remain invariant under the interchange of the $\tau$ and $x$ axes, which requires
$f_{\rm pot}(\vec{\theta}_\tau,\vec{\theta}_x;L_\tau,L_x) = f_{\rm pot}(\vec{\theta}_x,\vec{\theta}_\tau;L_x,L_\tau).
$
Moreover, in the limit $L_x\to\infty$, the model should reproduce the standard thermodynamics on $\mathbb{S}^1\times\mathbb{R}^3$, i.e.,
$
f_{\rm pot}(\vec{\theta}_\tau,\vec{\theta}_x;L_\tau,L_x) \rightarrow f_{\rm pot}^{\mathbb{S}^1\times\mathbb{R}^3}(\vec{\theta}_\tau,L_\tau).
$

The simplest potential satisfying these conditions is the “separable" form, expressed in terms of the eigenvalues of the two Polyakov loops:
\begin{align}
    f_{\rm sep}(\vec{\theta}_\tau,\vec{\theta}_x;L_\tau,L_x)=
    f^{\SRRR}_{\rm pot}(\vec{\theta}_\tau,L_\tau)
    + f^{\SRRR}_{\rm pot}(\vec{\theta}_x,L_x).
    \label{eq:fsep}
\end{align}
This construction, adopted in Ref.~\cite{Suenaga:2022rjk}, naturally reduces to the conventional thermodynamic model (model B) of Ref.~\cite{Meisinger:2001cq} as $L_x\to\infty$. 
Alternatively, $f_{\rm sep}(\vec{\theta}_\tau,\vec{\theta}_x;L_\tau,L_x)$ can be derived either by utilizing the model A from Ref.~\cite{Meisinger:2001cq} or the model presented in Ref.~\cite{Dumitru:2012fw}.  The model in Ref.~\cite{Dumitru:2012fw} extends model A to better fit lattice data over a wider $L_\tau$ region, and we adopt it in this work. In this model, the free energy is given by
\begin{align}
    f^{\SRRR}_{\rm pot}(\vec{\theta},L) 
    = -\frac{4\pi^2}{3}\frac{T_{\rm d}^2}{L^2}\Bigl(\frac{1}{5}c_1 V_1(\phi)+c_2 V_2(\phi)-\frac{2}{15}c_3\Bigr)
    +\frac{\tilde c_3}{2}\frac{8\pi^2T^4_{\rm d}}{45},
    \label{FDumitru}
\end{align}
where $V_1(\phi)$ and $V_2(\phi)$ are defined as
$V_1(\phi)= [2\phi(\pi-\phi)+\phi(2\pi-\phi)]/2\pi^2$ and 
$V_2(\phi)= [8\phi^2(\pi-\phi)^2+\phi^2(2\pi-\phi)^2]/8\pi^4$, 
the parameter set is $(c_1,c_2,c_3)=(0.552,\,0.830,\,0.950)$ with 
$\tilde c_3=(47-20c_2-27c_3)/27$. 
However, it turns out that the effective model combining the perturbative term (Eq.~\eqref{fpert2}) with the separable term (Eq.~\eqref{FDumitru}) does not fully capture the lattice results on $\mathbb{T}^2 \times \mathbb{R}^2$~\cite{Suenaga:2022rjk}. In particular, the lattice data indicate that the influence of the boundary conditions on the thermodynamic quantities remains negligible until $L_x$ becomes very small~\cite{Kitazawa:2019otp}, suggesting that an additional term is needed.

To address this shortcoming, we introduce cross terms, which allows for the interplay between the eigenvalues of the two Polyakov loops. As such a cross term, we assume the following form: 
\begin{align}
    f_{\rm cross}(\vec{\theta}_\tau,\vec{\theta}_x;L_\tau,L_x)&= \frac{T_{\rm d}^{4-2n}}{(L_\tau^2+L_x^2)^{n}} \Big[ c_4 X_\tau^{(1)}X_x^{(1)}
    + c_5 ( X_\tau^{(3)}X_x^{(1)} + X_\tau^{(1)}X_x^{(3)} )+ c_6 X_\tau^{(3)}X_x^{(3)} \Big],
    \label{eq:fcross}
\end{align}
with $X_c^{(1)}={\rm Tr}(\hat \Omega_c) {\rm Tr}(\hat \Omega_c^\dagger)$ and $X_c^{(3)}=\big[ {\rm Tr}(\hat \Omega_c^3) + {\rm Tr}(\hat \Omega_c^{\dagger3}) \big]/2$. This cross term should be written as a combination of $\Omega_c$ that is invariant under $Z^{(c)}_3$ transformations while satisfying the constraints discussed above. For simplicity, in this study, we consider all independent and possible combinations of $\Omega_c$ up to third order. 
We will demonstrate below that our effective model, constructed using this simple approach, successfully reproduces the lattice data on $\mathbb{T}^2\times\mathbb{R}^2$ with reasonable accuracy.

\subsection{Thermodynamics}
\label{sec:model:therm}

From the free energy introduced in the previous subsection, we define the thermodynamic quantities on $\mathbb{T}^2\times\mathbb{R}^2$. The boundary condition imposed along the $x$-axis breaks the rotational symmetry of the system, leading to anisotropic pressure. In particular, the pressure along the $x$-axis, $p_x$, differs from that along the $y$-axis, $p_y$, while, due to the rotational symmetry in the $y$-$z$ plane, we have $p_y=p_z$. Consequently, the energy-momentum tensor takes the diagonal form
$T^\mu_\nu(x)=\mathrm{diag}(\epsilon, p_x, p_z, p_z)$, 
and when $L_\tau=L_x$, the Euclidean directions $\tau$ and $x$ become equivalent, yielding $p_x=-\epsilon$.

Using the free energy defined in Eq.~\eqref{eq:f}, the thermodynamic quantities are given by
\begin{align}
    \epsilon = \frac{L_\tau}{\mathcal{V}}\frac{\partial}{\partial L_\tau}\Big(\mathcal{V}f\Big), \quad p_x = -\frac{L_x}{\mathcal{V}}\frac{\partial}{\partial L_x}\Big(\mathcal{V}f\Big), \quad p_y = p_z = -\frac{L_z}{\mathcal{V}}\frac{\partial}{\partial L_z}\Big(\mathcal{V}f\Big) = -f, \label{thermodynamics}
\end{align}
with the total volume defined as $\mathcal{V} = L_\tau L_x L_y L_z$. The interaction measure is defined by
$\Delta = T^\mu_\mu = \epsilon - p_x - 2p_z$, 
which vanishes when the system exhibits scale invariance. In this scale-invariant case, one finds that $p_x/p_z = -1$ for $L_\tau = L_x$.

\section{Results}

\subsection{Parameters}

Our model contains four free parameters, namely $c_4, c_5, c_6, n$, which are chosen to match the lattice thermodynamics data presented in Ref.~\cite{Kitazawa:2019otp}. By examining the dependence of $\TTRR$ on these parameters, we found that the following parameter set provides a good fit over a wide range of $L_\tau$ and $L_x$:
\begin{align}
    c_4=0.11, \ c_5=0.06, \ c_6=-0.03, \ n=1.85. \label{FCrossParameter}
\end{align}
Consequently, we adopt Eq.~\eqref{FCrossParameter} in the subsequent analysis. 
To the best of our knowledge, no parameter set can accurately reproduce the entire range of lattice data.

%%%%%%%%%%%%%%%%%%%%%%%%%%%%%%
\begin{figure*}
\includegraphics[width=0.48\textwidth]{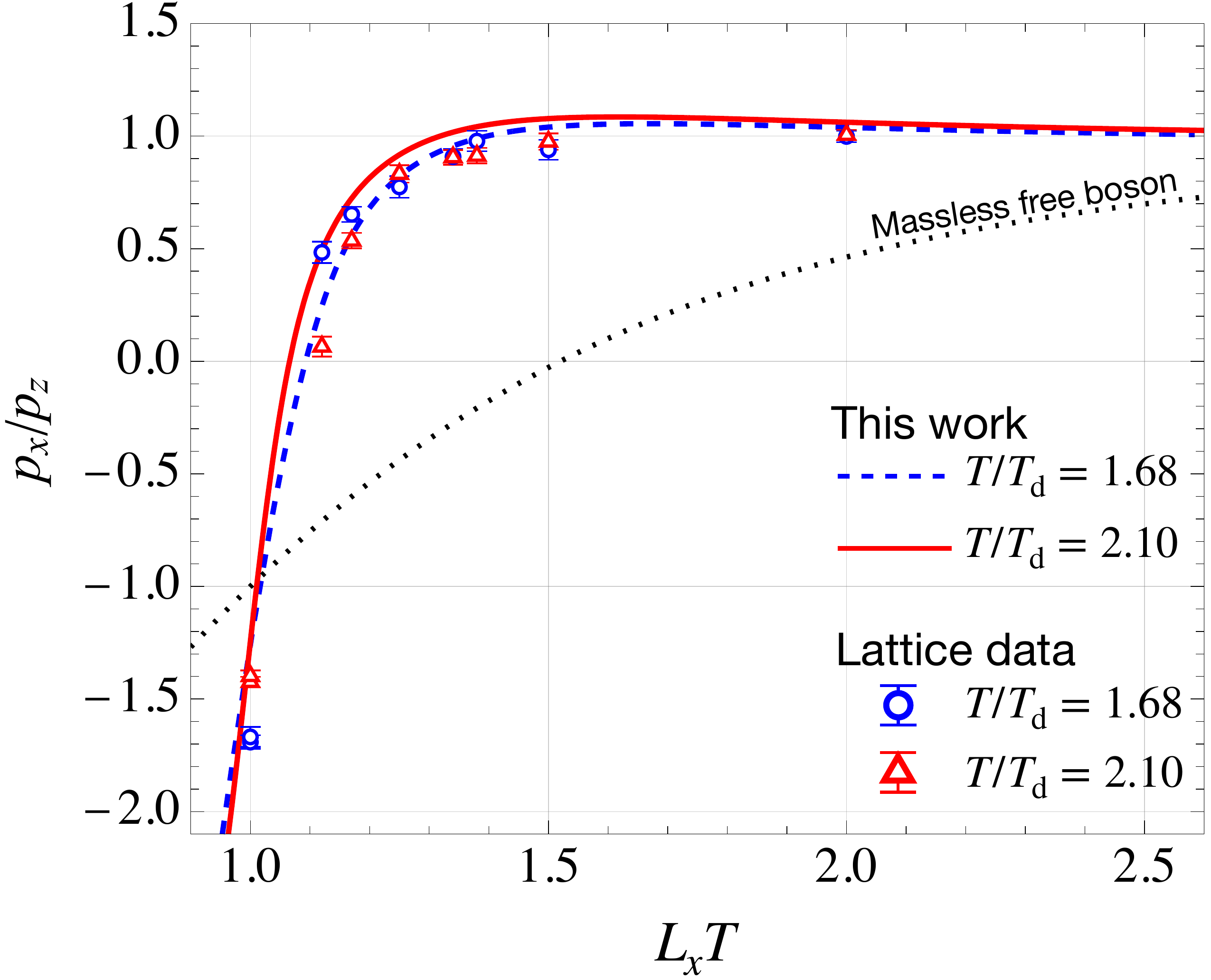}
\includegraphics[width=0.48\textwidth]{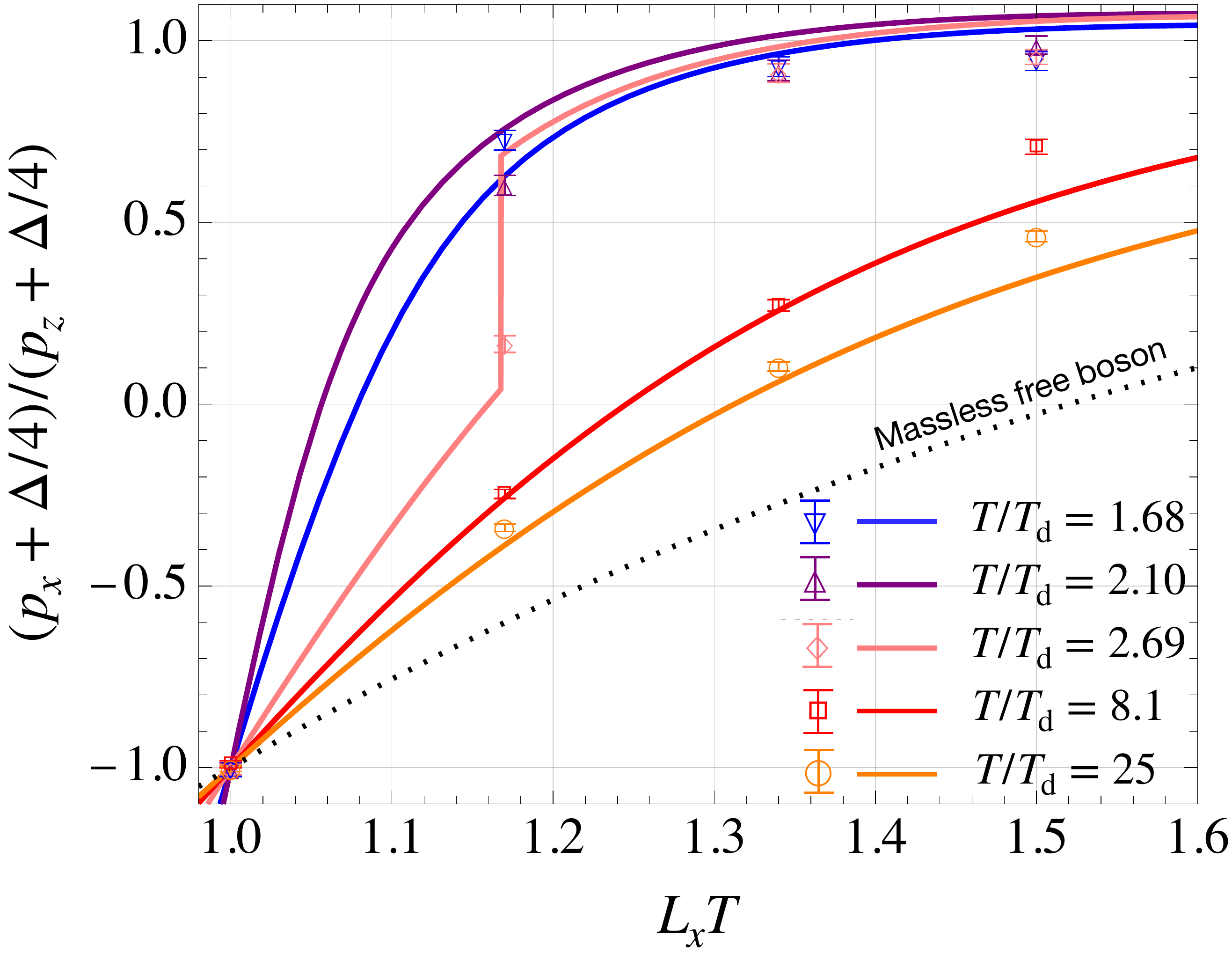}
\caption{\label{pxpz1} 
Left: Dependence of the ratio $p_x/p_z$ on $L_xT$ at $T/T_{\rm d}=1.68$ and $2.10$, together with lattice data from Ref.~\cite{Kitazawa:2019otp}. The dotted line represents the ratio for a massless free-boson system. 
Right: $L_xT$ dependence of the ratio $R=\frac{p_x+\Delta/4}{p_z+\Delta/4}$ for several values of $T/T_{\rm d}$. For details, see Ref.~\cite{Fujii:2024llh}}
\end{figure*}
%%%%%%%%%%%%%%%%%%%%%%%%%%%%%%

%%%%%%%%%%%%%%%%%%%%%%%%%%%%%%
\begin{figure}
\includegraphics[width=1\textwidth]{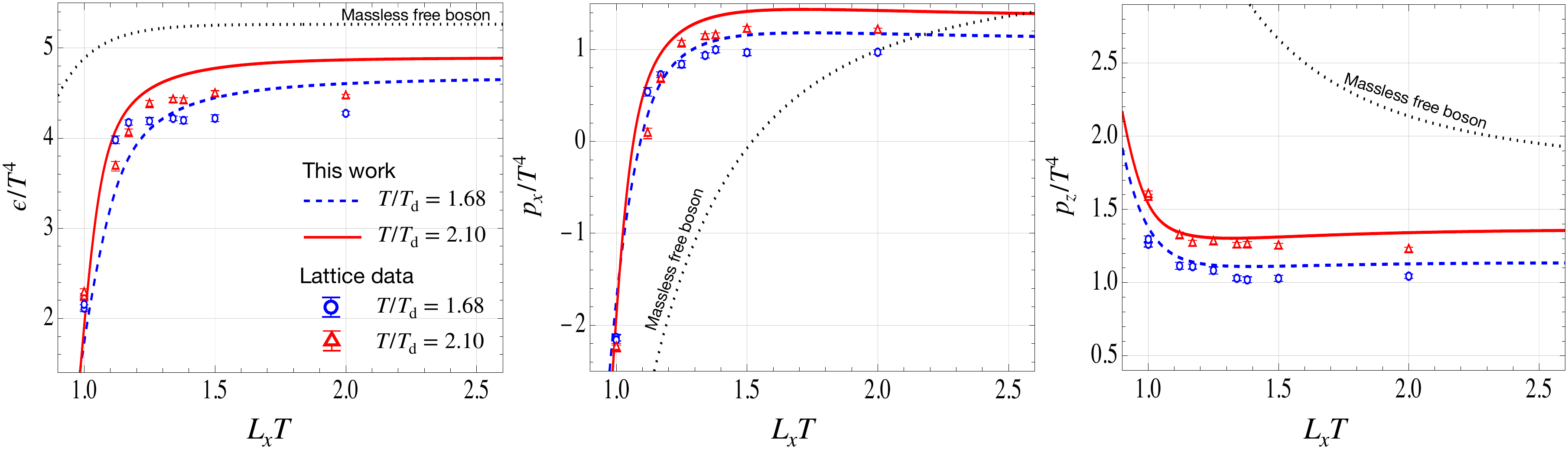}
\caption{\label{epxpz} 
$L_xT$ dependence of the energy density $\epsilon$ (left) and the pressures $p_x,\ p_z$ (middle and right) at $T/T_{\rm d}=1.68$ and $2.10$~\cite{Fujii:2024llh}. In all panels, the discrete markers represent lattice data from Ref.~\cite{Kitazawa:2019otp}.}
\end{figure}
%%%%%%%%%%%%%%%%%%%%%%%%%%%%%%

%%%%%%%%%%%%%%%%%%%%%%%%%%%%%%
\subsection{Thermodynamic quantities}
\label{sec:ThermodynamicsLattice}
%%%%%%%%%%%%%%%%%%%%%%%%%%%%%%

We compare the thermodynamic quantities defined in Eq.~(\ref{thermodynamics}) from our effective model with lattice data for $\mathbb{T}^2\times\mathbb{R}^2$. 
In the left panel of Fig.~\ref{pxpz1}, we display the $L_xT$ dependence of the ratio $p_x/p_z$ at $T/T_{\rm d}=1.68$ (blue dashed line) and $2.10$ (red solid line), together with lattice results from Ref.~\cite{Kitazawa:2019otp} (shown as discrete points with error bars). In an isotropic system, one expects $p_x/p_z=1$, i.e., $\lim_{L_xT\to\infty}p_x/p_z=1$, and any deviation from unity serves as a measure of anisotropy. In the figure, the dotted line shows the behavior of $p_x/p_z$ for a massless free-boson system, which deviates noticeably from unity already at $L_xT=2.5$. In contrast, the lattice data remain near unity down to $L_xT\approx1.3$, before abruptly dropping to negative values for $L_xT\lesssim1.2$.
These observations indicate that our model, which incorporates a cross term, qualitatively reproduces the lattice results. This stands in contrast to the earlier model without the cross term reported in Ref.~\cite{Suenaga:2022rjk}, highlighting the crucial role played by the interplay between $\Omega_\tau$ and $\Omega_x$ %on $\mathbb{T}^2\times\mathbb{R}^2$.

At high temperatures, analysis becomes more challenging. To avoid technical issues in this regime, we adopt the ratio $R = \frac{p_x+\Delta/4}{p_z+\Delta/4}$, 
as a thermodynamic observable according to Ref.~\cite{Kitazawa:2019otp}. In the right panel of Fig.~\ref{pxpz1}, we compare our model predictions with lattice data for temperatures up to \(T/T_{\rm d}\simeq25\); here, \(R\) is plotted as a function of \(L_xT\) for various values of \(T/T_{\rm d}\). Note that, since \(p_x=-\epsilon\), the condition at \(L_xT=1\) implies \(R=-1\), a result that is confirmed by both the model and the lattice data.

Further inspection of the right panel shows that the behavior of $R$ differs markedly between low and high temperatures. For $T/T_{\rm d}=1.68$ and $2.10$, $R$ undergoes a sharp change around $L_xT\simeq1.2$, consistent with the observations in the left panel. Conversely, at $T/T_{\rm d}=8.1$ and $25$, $R$ varies more smoothly. Notably, at $T/T_{\rm d}=2.69$, while the behavior for $L_xT\gtrsim1.3$ resembles that at lower temperatures, a distinct drop is observed at $L_xT\approx7/6\simeq1.17$, indicating a discontinuous jump corresponding to a first-order phase transition. 
This first-order phase transition will be discussed in detail in the next subsection, showing the phase diagram and the behavior of the Polyakov loops. 

Finally, Fig.~\ref{epxpz} shows the $L_xT$ dependence of the energy density $\epsilon$ and the pressures $p_x$ and $p_z$ at $T/T_{\rm d}=1.68$ and $2.10$. The model predictions (depicted as continuous lines) qualitatively match the lattice data, although small differences become apparent at larger $L_xT$. These differences mainly arise from the thermodynamic model $f_{\rm pot}^{\SRRR}$ as presented in Ref.~\cite{Dumitru:2012fw}, which implies that we do not expect perfect quantitative agreement in this range.

%%%%%%%%%%%%%%%%%%%%%%%%%%%%
\subsection{Phase diagram}
\label{sec:PhaseStructure}
%%%%%%%%%%%%%%%%%%%%%%%%%%%%

%%%%%%%%%%%%%%%%%%%%%%%%%%%%%%
\begin{figure*}
\includegraphics[width=0.45\textwidth]{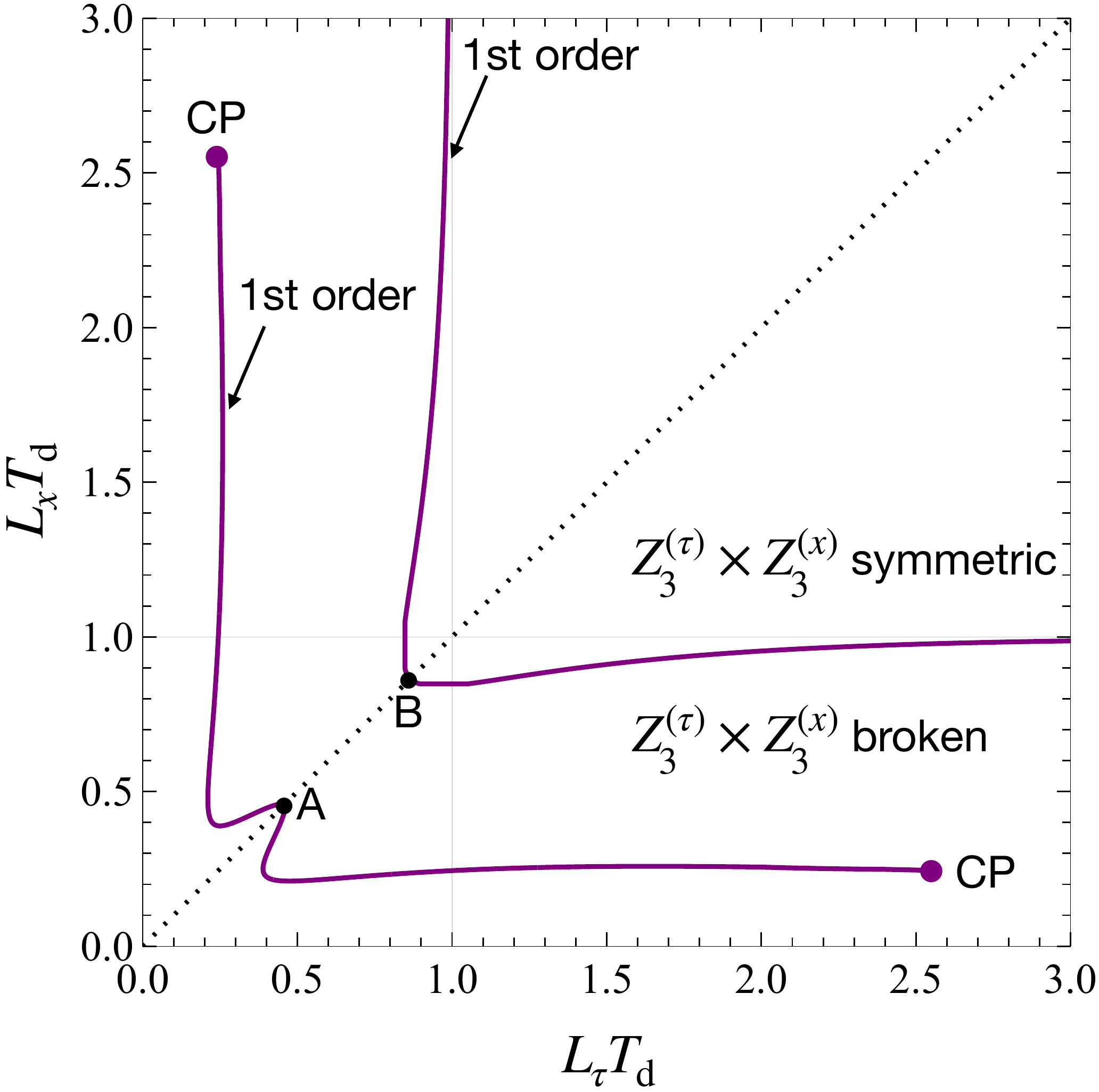}
\includegraphics[width=0.45\textwidth]{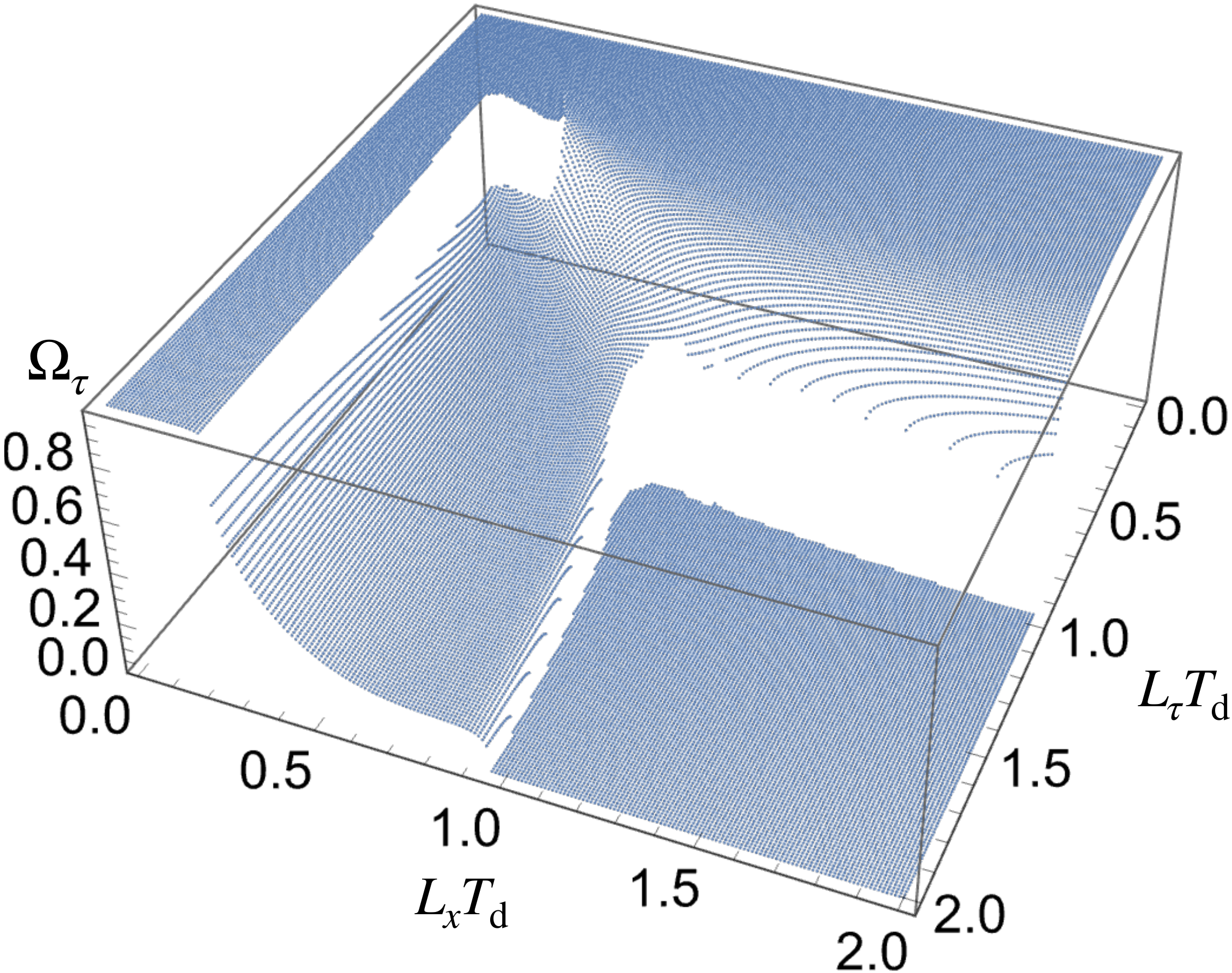}
\caption{\label{fig:PhaseDiagram} 
Left: Phase diagram in the $L_\tau$–$L_x$ plane. Solid lines mark the first-order transitions where both the thermodynamic quantities and the Polyakov loop variable $\Omega_c$ change discontinuously, with critical points (CPs) highlighted by large circles. 
Right: Behavior of $\Omega_\tau$ on the $L_\tau$–$L_x$ plane. (Note: the corresponding behavior of $\Omega_x$ can be obtained by interchanging the $\tau$ and $x$ axes.) For details, see Ref~\cite{Fujii:2024llh}.
 }
\end{figure*}
%%%%%%%%%%%%%%%%%%%%%%%%%%%%%%

The left panel of Fig.~\ref{fig:PhaseDiagram} presents the phase diagram of our model on the \(L_\tau\)–\(L_x\) plane. Here, solid lines indicate first-order phase transitions where the thermodynamic quantities change discontinuously, and the dotted line along \(L_\tau = L_x\) reflects the symmetry under exchanging the \(\tau\) and \(x\) directions. In the right panel, we show the behavior of the Polyakov loop variables \(\Omega_\tau\) (top) and \(\Omega_x\) (bottom) over the same plane.

These figures reveal two distinct first-order transition lines in the \(L_\tau\)–\(L_x\) space, where both \(\Omega_\tau\) and \(\Omega_x\) exhibit discontinuous jumps concomitant with changes in the thermodynamic quantities. One of these lines, which passes through point B in the left panel, connects to the confined transition on \(\SRRR\) in the limit of large \(L_x\) (or \(L_\tau\)). In the region above and to the right of this line, the system is in the confined phase with both \(Z_3^{(c)}\) symmetries restored (\(\Omega_\tau = \Omega_x = 0\)); conversely, in the lower-left region, both \(Z_3^{(c)}\) symmetries are spontaneously broken (\(\Omega_\tau \neq 0\) and \(\Omega_x \neq 0\)). Notably, our analysis indicates that there is no phase in which only one of the \(Z_3^{(c)}\) symmetries is broken.

The second first-order transition line, which includes point A, corresponds to the transition observed in the right panel of Fig.~\ref{pxpz1}. As shown in Fig.~\ref{fig:PhaseDiagram}, this line lies entirely within the \(Z_3^{(\tau)} \times Z_3^{(x)}\) broken phase and terminates at finite values of \(L_\tau\) and \(L_x\) at approximately \((L_\tau T_{\rm d}, L_x T_{\rm d}) \simeq (2.54, 0.25)\) and \((0.25, 2.54)\). This line is disconnected from any transition on \(\SRRR\). Its endpoint corresponds to a critical point (CP) where the first-order transition becomes second order. The critical behavior at this CP falls into the universality class of the two-dimensional Ising model (\(Z_2\) universality class).

\section{Summary}

In this proceedings, we investigated the thermodynamics of $SU(3)$ Yang-Mills theory on $\mathbb{T}^2\times\mathbb{R}^2$ using an effective model that includes two dynamical Polyakov loops, $\Omega_\tau$ and $\Omega_x$, extended with cross terms to capture their interplay. Our model successfully reproduces the qualitative lattice data of Ref.~\cite{Kitazawa:2019otp}. 
Notably, we predict a novel first-order phase transition within the \(Z_3^{(\tau)}\times Z_3^{(x)}\) broken phase—where both \(\Omega_\tau\) and \(\Omega_x\) remain nonzero—that is distinct from the deconfined transition and terminates at potential critical endpoints belonging to the two-dimensional \(Z_2\) universality class. 
These findings motivate further lattice studies with improved resolution and suggest promising avenues for extending the analysis to other theories~\cite{Hanada:2010kt,Unsal:2010qh,Mandal:2011hb,GarciaPerez:2014cmv,Chernodub:2018aix,Chernodub:2022izt,Hayashi:2024qkm,Fujii:2024fzy,Fujii:2024ixq,Fujii:2024woy} and deriving the effective potential from first-principles approaches.

%%%%%%%%%%%%%%%%%%%
\section*{Acknowledgments}
%%%%%%%%%%%%%%%%%%%

This work was supported in part by JSPS KAKENHI (Nos. JP22K03619, JP23K03377, JP23H04507, JP23H05439, JP24K07049, JP24K17054),
the Center for Gravitational Physics and Quantum Information (CGPQI) at Yukawa Institute for
Theoretical Physics.

\bibliographystyle{JHEP}
\bibliography{reference}

%\begin{thebibliography}{99}
%\bibitem{...}
%....

%\end{thebibliography}

\end{document}